# Multiple-Carrier-Lifetime Model for Carrier Dynamics in InGaN/GaN LEDs with Non-Uniform Carrier Distribution

Xuefeng Li[1*], Elizabeth DeJong[1], Rob Armitage[2], and Daniel Feezell[1]

[1]*Center for High Technology Materials (CHTM), University of New Mexico, Albuquerque, NM 87106, USA*
[2]*Lumileds LLC, San Jose, CA 95131, USA*

We introduce a multiple-carrier-lifetime model (MCLM) for light-emitting diodes (LEDs) with non-uniform carrier distribution, such as in multiple-quantum-well (MQW) structures. By employing the MCLM, we successfully explain the modulation response of V-pit engineered MQW LEDs, which exhibit an $S_{21}$ roll-off slower than -20 dB/decade. Using the proposed model and employing a gradient descent method, we extract effective recombination and escape lifetimes by averaging the carrier behavior across the quantum wells. Our results reveal slower effective carrier recombination and escape in MQW LEDs compared with LEDs emitting from a single QW, indicating the advantages of lower carrier density achieved through V-pit engineering. Notably, the effective carrier recombination time is more than one order of magnitude lower than the effective escape lifetime, suggesting that most carriers in the quantum wells recombine, while the escape process remains weak. To ensure the reliability and robustness of the MCLM, we subject it to a comprehensive three-fold validation process. This work confirms the positive impact of spreading carriers into several QWs through V-pit engineering. In addition, the MCLM is applicable to other LEDs with non-uniform carrier distribution, such as micro-LEDs with significant surface recombination and non-uniform lateral carrier profiles.

___________________________________
[*]**Electronic mail:** xuefengli@unm.edu

Achieving highly efficient RGB solid-state lighting remains a challenge due to the low internal quantum efficiency (IQE) in green LEDs, which is known as the "green gap" [1]. Research has identified several potential factors contributing to this, including Auger-Meitner recombination [2-5], the quantum-confined Stark effect (QCSE) [6-9], the carrier-current density relationship [10], carrier leakage [11,12], and defects and carrier localization [1,13-18]. Increasing the number of quantum wells (QWs) is a common approach to addressing the green gap due to a larger QW region volume, which is beneficial for reducing the carrier density at a given current density and hence decreasing the intrinsic Auger-Meitner recombination. In our previous study, we identified intrinsic Auger-Meitner recombination as the primary nonradiative recombination mechanism contributing to efficiency loss in green LEDs using state-of-the-art growth conditions [19]. However, achieving uniform carrier distribution across the QWs is difficult in green InGaN/GaN LEDs. Due to strong hole confinement and slow transport in MQWs [20,21], the carriers in one-dimensional device designs are mostly confined to the QW next to the p-GaN, as observed both in simulation [22,23] and experiment [24,25]. Techniques such as V-pit engineering (three-dimensional device designs) can improve carrier transport and increase the effective volume of the active region but still do not achieve an entirely uniform carrier distribution among the MQW [26-30]. It's important to understand the carrier dynamics in InGaN/GaN MQW LEDs to further improve the quantum efficiency. To characterize the carrier dynamics in InGaN/GaN LEDs, a single-carrier-lifetime model (SCLM) is usually used to analyze the carrier behavior in the active region using small-signal electroluminescence (SSEL) methods [31,32]. However, problems arise when applying the SCLM to MQW LEDs with non-uniform carrier distribution. For example, the modulation response of an MQW LED cannot be fit with the SCLM, which requires an $S_{21}$ roll-off close to -20 dB/decade [31,32]. The fundamental reason behind this limitation is that different regions within the MQWs have different carrier lifetimes due to the differences in potential energy, internal polarization, and carrier distribution across the active region. Developing an accurate carrier lifetime model for InGaN/GaN MQW LEDs with non-uniform carrier distribution will be beneficial for optimizing their quantum efficiency. Such a model will be described by multiple carrier lifetimes.

In this work, we present a multiple-carrier-lifetime model (MCLM) for InGaN/GaN LEDs using SSEL [31,32]. Instead of assuming the carriers recombine at the same rate over the entire active region, we decouple the QWs into multiple regions, each with different carrier lifetimes. The small-signal equivalent circuit is obtained, and a gradient descent method is used to acquire the circuit elements related to carrier transportation and recombination. Then, effective differential recombination and escape lifetimes are obtained at various current densities. The new model is capable of analyzing LEDs with $S_{21}$ roll-off that is slower than -20 dB/decade, which is impossible with the SCLM. Furthermore, we conduct a three-fold validation process on the MCLM to confirm the reliability and robustness of the model.

The SCLM and MCLM are applied to two LED structures in this work. The first structure has 4 X 3 nm QWs with only one QW emitting, and the other structure has 12 X 3 nm QWs with V-pit engineering and multiple QWs emitting. The two structures are hereafter referred to as "quasi-SQW LED" and "MQW LED," respectively. The indium content in the QWs of both wafers is 19% based on energy-dispersive x-ray spectroscopy (EDX) measurements, and the GaN barriers are 18 nm thick. In the quasi-SQW LED, carrier recombination is confined to a single QW, but multiple QWs contribute to photon emission in the MQW LED with V-pit engineering.

The internal quantum efficiency (IQE), shown in Figure 1(a), was obtained from Lumileds by measuring the external quantum efficiency (EQE) of LEDs with known extraction efficiency. The IQE in the MQW LED with V-pit engineering was higher than that of the quasi-SQW LED at all current density ($J$) values studied, especially at high $J$. The IQE increased from 27.1% in the quasi-SQW LED, to 36.6% in the MQW LED at 40 A/cm$^2$. The -3dB bandwidths vs. current density for the two LED are shown in Figure 1(b). The -3dB bandwidths were obtained through small-signal S-parameter measurements and a detailed discussion can be found in Ref. 33. Both bandwidths increase at high $J$ and the -3dB bandwidth of the quasi-SQW LED is several times higher than that of the MQW LED. The -3dB bandwidths are 20.2 MHz and 4.2 MHz at 40 A/cm$^2$ for the quasi-SQW and MQW LEDs, respectively.

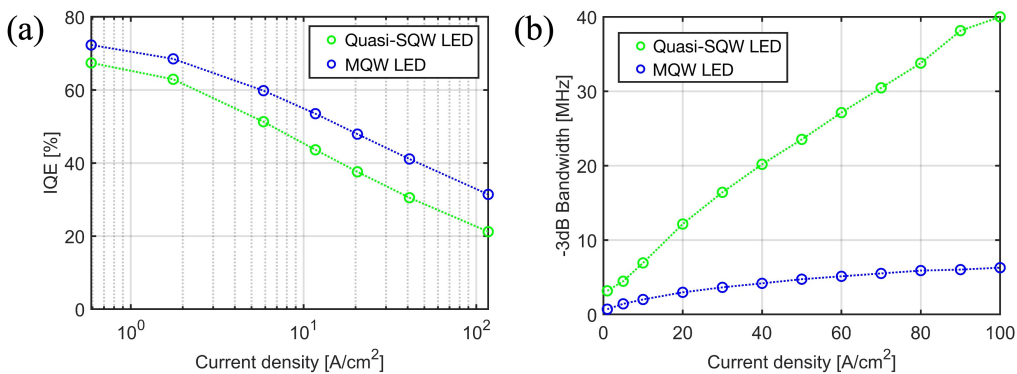

Figure 1. (a) IQE and (b) -3dB bandwidth for quasi-SQW and MQW LEDs.

Next, we introduce the conventional SCLM and investigate the $S_{21}$ roll-off. Figure 2 illustrates the equivalent circuit of the SCLM. More comprehensive descriptions can be found in Ref. 31 and Ref. 32. Here, $R_w$ and $C_w$ are resistance and capacitance associated with the QW region, $R_c$ is the resistance associated with carrier transport in the cladding layer, $R_{rc}$ is the resistance associated with carrier recombination in the cladding layer, $C_t$ is the capacitance associated with space-charge in the PIN junction and free carriers in the cladding layer, and $R_s$ is the parasitic resistance associated with the contacts. $V_{in}, V_{ref}$, and $V_{out}$ are input, reference, and output voltage in the equivalent circuit, respectively.

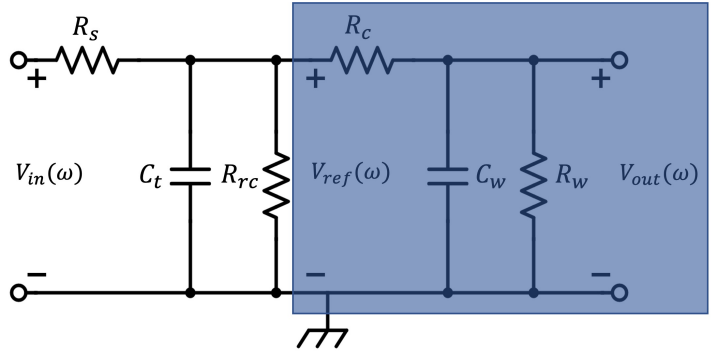

Figure 2. Equivalent circuit of the SCLM.

Figures 3(a) and 3(b) show the measured impedance and modulation responses fit with the SCLM for quasi-SQW and MQW LEDs with 100-µm-diameter injection areas. We observed a higher impedance and lower modulation bandwidth in the MQW LED due to the better carrier transport and lower carrier density in the QWs. Furthermore, the S$_{21}$ roll-off in the MQW LED is around -18 dB/decade, while the S$_{21}$ roll-off in the quasi-SQW LED is -20 dB/decade. The fitting of S$_{21}$ in the MQW LED is inferior to the fitting of S$_{21}$ in the quasi-SQW LED using the SCLM. In the next section, we provide a theoretical explanation for why the SCLM cannot account for or properly model S$_{21}$ roll-offs that are slower than -20 dB/decade.

Based on the equivalent circuit shown in Figure 2, the modulation response of the SCLM can be expressed as:

$$\left|\frac{V_{out}}{V_{in}}\right|^2 = \left|\frac{V_{out}}{V_{ref}} \times \frac{V_{ref}}{V_{in}}\right|^2 = \left|\frac{V_{out}}{V_{ref}}\right|^2 \times \left|\frac{Z_{tot} - R_s}{Z_{tot}}\right|^2 \quad (1)$$

The second term on the right-hand side is the cladding related modulation response and is only related to the total impedance ($Z_{tot}$) and $R_s$. After converting the cladding related modulation response into decibels (dB) and normalizing it with respect to the modulation response at low frequencies, it was found that the cladding-related normalized power (second term on the right-hand side of equation (1)) is below -0.4 dB/decade within the studied operational frequency range, as depicted in Figure 4. Thus, the impact of the cladding-related modulation response on the overall modulation response is minimal. On the other hand, the QW related modulation response, represented by the first term on the right-hand side of equation (1), is dominant and can be expressed as:

$$\left|\frac{V_{out}}{V_{ref}}\right|^2 = \frac{R_w^2}{(R_c + R_w)^2 + R_c^2 R_w^2 C_w^2 \omega^2} = \frac{1}{(\frac{R_c}{R_w} + 1)^2 + (\frac{R_c}{R_w})^2 \omega^2 \Delta\tau_{rec}^2} \quad (2)$$

Here, $\Delta\tau_{rec} = R_w C_w$ is the differential recombination lifetime in the QW active region. After converting the modulation response into dB, equation (2) can be formulated as:

$$10\log_{10}\left(\left|\frac{V_{out}}{V_{in}}\right|^2\right) = -10\log_{10}\left(1 + \omega^2 \Delta\tau_{rec}^2 \frac{1}{\left(1 + \frac{R_w}{R_c}\right)^2}\right) + 10\log_{10}\left(\frac{1}{\left(1 + \frac{R_c}{R_w}\right)^2}\right) \quad (3)$$

The second term on the right is a constant value for a given circuit and current density, and it's irrelevant to the roll-off of the modulation response. At high frequencies, $\omega^2 \Delta\tau_{rec}^2 \frac{1}{\left(1 + \frac{R_w}{R_c}\right)^2} \gg 1$. Otherwise, no decline in modulation would be observed. Thus, equation (3) can be reformulated as:

$$10\log_{10}\left(\left|\frac{V_{out}}{V_{in}}\right|^2\right) = -10\log_{10}\left(\omega^2 \Delta\tau_{rec}^2 \frac{1}{\left(1+\frac{R_w}{R_c}\right)^2}\right) + 10\log_{10}\left(\frac{1}{\left(1+\frac{R_c}{R_w}\right)^2}\right)$$

$$= -20\log_{10}(\omega) - 10\log_{10}\left(\frac{\Delta\tau_{rec}^2}{\left(1+\frac{R_w}{R_c}\right)^2}\right) + 10\log_{10}\left(\frac{1}{\left(1+\frac{R_c}{R_w}\right)^2}\right) \quad (4)$$

The $S_{21}$ roll-off is solely determined by the first term, while the second and third terms remain constant regardless of the operating frequency. The $S_{21}$ roll-off is defined as the reduction in modulation response as the frequency increases by a decade, and can be expressed as:

$$-20\log_{10}\left(\frac{\omega_2}{\omega_1}\right) = -20 \; dB/decade \quad (5)$$

Here, $\omega_2 = 10 \times \omega_1$. Therefore, the SCLM can only predict and model an $S_{21}$ roll-off of -20 dB/decade. Thus, LEDs with non-uniform carrier distribution and roll-offs other than -20 dB/decade require a modified model.

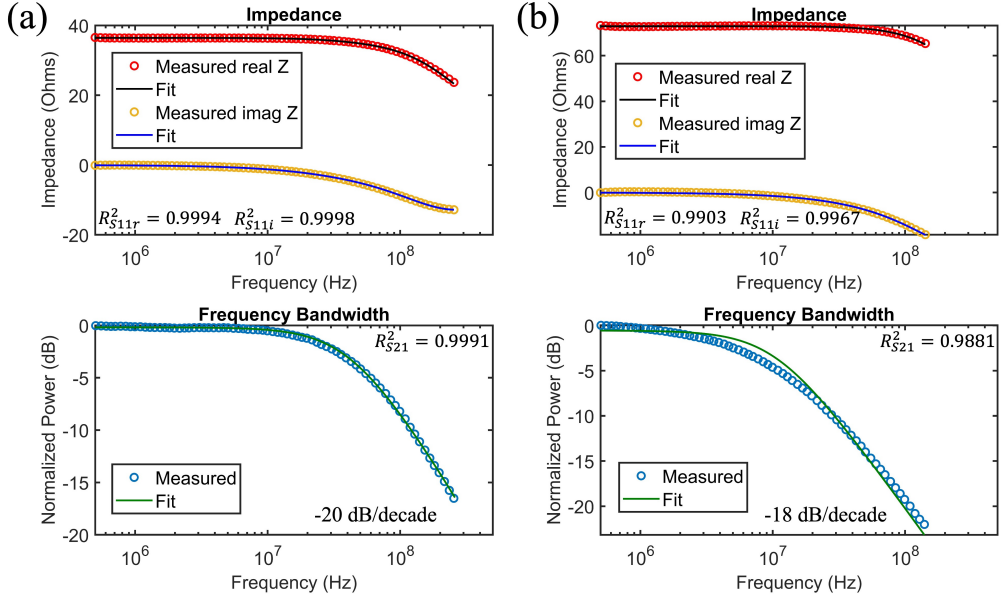

Figure 3. Fitting of impedance and modulation response of (a) quasi-SQW and (b) MQW LEDs using the SCLM.

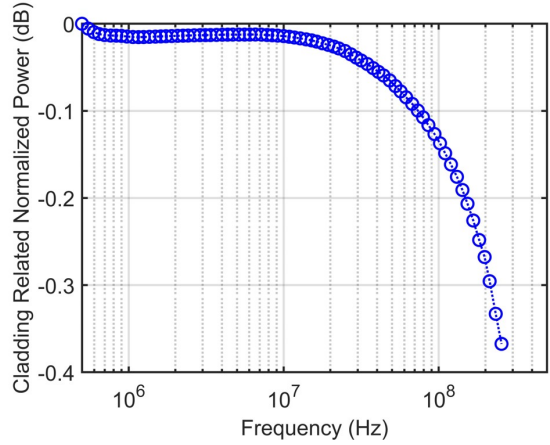

Figure 4. Cladding-related normalized power.

Compared with the uniform carrier distribution model, the small-signal equivalent circuit for an LED with non-uniform carrier distribution includes several QW regions, each with different carrier recombination lifetimes. Here, we assume random QW regions with an index of $j$. All the parameters below with a subscript $j$ are the corresponding parameters of QW region $j$. Rate equations capture the carrier dynamics in both the active region and cladding layers. There are several mechanisms considered in the rate equations: carrier injection $\frac{I}{q}$, carriers charging the space-charge capacitance of the junction $\frac{C_{sc}}{q}\frac{dV_c}{dt}$, carriers captured by the QWs $\frac{N_c}{\tau_{cj}}$, carriers recombined in the QWs $\frac{N_{wj}}{\tau_{recj}}$, and carriers escaped $\frac{N_{wj}}{\tau_{escj}}$ from the QWs. Here, $q$ is elementary charge, $C_{sc}$ is space charge capacitance, $V_c$ is the junction voltage, $N_c$ is the population of carriers in the cladding layer, $N_{wj}$ is the population of carriers, $\tau_{cj}$ is carrier transport lifetime, $\tau_{recj}$ is carrier recombination lifetime, and $\tau_{escj}$ is carrier escape lifetime, all in QW region $j$, respectively. In this analysis, we ignore the carrier recombination in the cladding layer as the process has limited effect on the differential carrier lifetime analysis in SSEL [32,34]. Therefore, the differential rate equations can be formulated as follows:

$$\frac{d}{dt}(dN_c) = \frac{dI}{q} - \frac{C_{sc}}{q}\frac{d}{dt}(dV_c) + \sum_{j=1}^{n}\frac{1}{\Delta\tau_{escj}}(dN_{wj}) - \sum_{j=1}^{n}\frac{1}{\Delta\tau_{cj}}(dN_c) \tag{6}$$

$$\frac{d}{dt}(dN_{w1}) = -\left(\frac{1}{\Delta\tau_{rec1}} + \frac{1}{\Delta\tau_{esc1}}\right)dN_{w1} + \frac{dN_c}{\Delta\tau_{c1}} \tag{7}$$

$$\frac{d}{dt}(dN_{w2}) = -\left(\frac{1}{\Delta\tau_{rec2}} + \frac{1}{\Delta\tau_{esc2}}\right)dN_{w2} + \frac{dN_c}{\Delta\tau_{c2}} \tag{8}$$

$$\vdots$$

$$\frac{d}{dt}(dN_{wj}) = -\left(\frac{1}{\Delta\tau_{recj}} + \frac{1}{\Delta\tau_{escj}}\right)dN_{wj} + \frac{dN_c}{\Delta\tau_{cj}} \tag{9}$$

After a small signal analysis similar to Ref. 32, the matrix format of the differential rate equations can be expressed as:

$$\begin{pmatrix} j\omega(C_c + C_{sc}) + \sum_{j=1}^{n}\frac{1}{R_{cj}} & -\frac{1}{R_{c1}} & -\frac{1}{R_{c2}} & \cdots & -\frac{1}{R_{cj}} & \cdots \\ -\frac{1}{R_{c1}} & j\omega C_{w1} + \frac{1}{R_{w1}} + \frac{1}{R_{c1}} & 0 & \cdots & 0 & \cdots \\ -\frac{1}{R_{c2}} & 0 & j\omega C_{w2} + \frac{1}{R_{w2}} + \frac{1}{R_{c2}} & \cdots & 0 & \cdots \\ \vdots & \vdots & \vdots & \ddots & 0 & \cdots \\ -\frac{1}{R_{cj}} & 0 & 0 & 0 & j\omega C_{wj} + \frac{1}{R_{wj}} + \frac{1}{R_{cj}} & \cdots \\ \vdots & \vdots & \vdots & \vdots & \vdots & \ddots \end{pmatrix} \times \begin{pmatrix} v_c(\omega) \\ v_{w1}(\omega) \\ v_{w2}(\omega) \\ \cdots \\ v_{wj}(\omega) \\ \cdots \end{pmatrix} = \begin{pmatrix} i(\omega) \\ 0 \\ 0 \\ \cdots \\ 0 \\ \cdots \end{pmatrix} \tag{10}$$

Here, $R_{cj} = \frac{\Delta\tau_{cj}}{C_c}$, and $R_{wj} = \frac{\Delta\tau_{recj}}{C_{wj}}$, are the resistances associated with carriers in the cladding layer and QW region, respectively. The differential escape lifetime in the QW region is: $\Delta\tau_{escj} = R_{cj}C_{wj}$. The total capacitance in the cladding region is: $C_t = C_c + C_{sc}$. The matrix in equation (10) corresponds to the small-signal equivalent circuit shown in Figure 5. A parasitic resistance $R_s$ is included to account for the resistance in the cladding layer and contact.

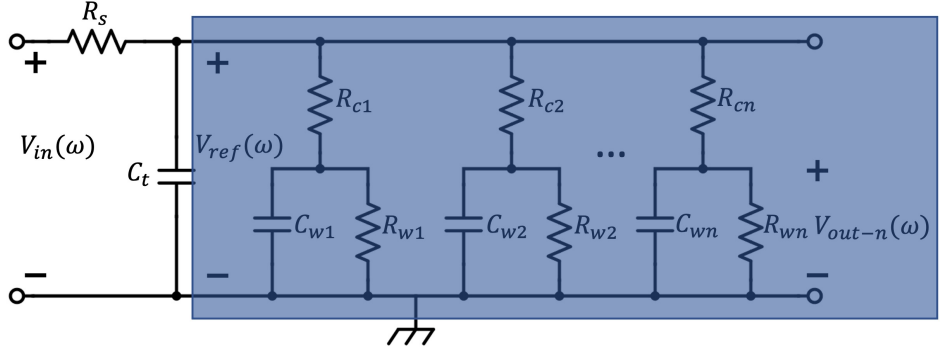

Figure 5. Equivalent circuit of the MCLM.

The differential rate equation for QW region $j$ was described in equation (9). At steady state, carriers injected into QW region $j$ will recombine or escape:

$$\frac{dN_c}{R_{cj}C_c} = \frac{dN_{wj}}{R_{wj}C_{wj}} + \frac{dN_{wj}}{R_{cj}C_{wj}} \tag{11}$$

Then, the ratio between the carrier population in QW region $j$ and QW region 1 is:

$$\frac{dN_{wj}}{dN_{w1}} = \frac{R_{c1}}{R_{cj}} \times \frac{(\frac{1}{R_{w1}} + \frac{1}{R_{c1}})C_{wj}}{(\frac{1}{R_{wj}} + \frac{1}{R_{cj}})C_{w1}} \tag{12}$$

Hence, the ratio of the corresponding carrier recombination rates is:

$$\frac{\frac{dN_{wj}}{R_{wj}C_{wj}}}{\frac{dN_{w1}}{R_{w1}C_{w1}}} = \frac{dN_{wj}R_{w1}C_{w1}}{dN_{w1}R_{wj}C_{wj}} = \frac{R_{w1}R_{c1}}{R_{wj}R_{cj}} \times \frac{\left(\frac{1}{R_{w1}} + \frac{1}{R_{c1}}\right)}{\left(\frac{1}{R_{wj}} + \frac{1}{R_{cj}}\right)} = \frac{R_{w1} + R_{c1}}{R_{wj} + R_{cj}} \tag{13}$$

We notice that the carrier recombination rate in QW region $j$ is inversely proportional to the total impedance of the QW region $j$ (i.e., $R_{wj} + R_{cj}$) at zero frequency, which matches with the LED equivalent circuit analysis in Figure 5.

The modulation response of the LED describes light output corresponding to the change of the drive frequency and is the sum of the modulation response of each QW region. The modulation response of QW region $j$ is:

$$\left|\frac{V_{out-j}}{V_{in}}\right|^2 = \left|\frac{Z_{QWj}}{Z_j} \times \frac{Z_{tot} - R_s}{Z_{tot}}\right|^2 \tag{13}$$

Here, $Z_{QWj}$ is the impedance of $R_{wj}$ and $C_{wj}$ elements in parallel, $Z_j$ is the impedance of $Z_{QWj}$ and $R_{cj}$ elements in series, $Z_{tot}$ is the total impedance of the LED equivalent circuit. Taking the corresponding carrier recombination rate described above into account, the modulation response of QW region $j$ can be rewritten as: $k \times \frac{1}{R_{wj}+R_{cj}} \times \left|\frac{Z_{QWj}}{Z_j} \times \frac{Z_{tot}-R_s}{Z_{tot}}\right|^2$. Here, $k$ is a constant with unit of Ohms.

Then, the modulation response of the whole MQW active region is:

$$k \times \sum_{j=1}^{n} \frac{1}{R_{wj} + R_{cj}} \times \left|\frac{Z_{QWj}}{Z_j} * \frac{Z_{tot} - R_s}{Z_{tot}}\right|^2 = k \times \left|\frac{Z_{tot} - R_s}{Z_{tot}}\right|^2 \times \sum_{j=1}^{n} \frac{1}{R_{wj} + R_{cj}} \times \left|\frac{Z_{QWj}}{Z_j}\right|^2 \tag{14}$$

Here, the term $\left|\frac{Z_{tot}-R_s}{Z_{tot}}\right|^2$ has very limited effect on the modulation response of each QW as described in the previous section and can be set to 1 in the analysis. The modulation response of any given QW region $j$ is similar to that of the SCLM and should also generate an S$_{21}$ roll-off of -20 dB/decade. The modulation response of the whole QW active region is the sum of all the QW regions and can be expressed similarly to equation (2) as:

$$k \times \sum_{j=1}^{n} \frac{1}{R_{wj} + R_{cj}} \times \left|\frac{Z_{QWj}}{Z_j}\right|^2 = k \times \sum_{j=1}^{n} \frac{1}{R_{wj} + R_{cj}} \times \frac{R_{wj}^2}{(R_{wj} + R_{cj})^2 + R_{cj}^2 R_{wj}^2 C_{wj}^2 \omega^2}$$

$$= k \times \sum_{j=1}^{n} \frac{r_j}{R_{wj}(r_j + 1)} \times \frac{r_j^2}{(r_j + 1)^2 + \Delta \tau_{recj}^2 \omega^2} \tag{15}$$

Here, $\Delta \tau_{recj} = R_{wj} C_{wj}$ is associated with the differential carrier lifetime in QW region $j$, and $r_j = \frac{R_{wj}}{R_{cj}}$ is the impedance ratio between $R_{wj}$ and $R_{cj}$, and is associated with emission intensity. Equation (15) shows that a modulation response with an S$_{21}$ roll-off slower than -20 dB/decade can be generated by multiple modulation responses with S$_{21}$ roll-offs of -20 dB/decade but with different emission intensities. The details will be presented below.

In contrast to the SCLM, the MCLM allows for different lifetimes from different QW regions. To get a representative recombination lifetime for the whole active region, we define the effective differential carrier recombination lifetime $\Delta\tau_{rec-eff}$:

$$\sum_{j=1}^{n} \frac{dN_{wj}}{\Delta\tau_{recj}} = \frac{1}{\Delta\tau_{rec-eff}} \sum_{j=1}^{n} dN_{wj} \tag{16}$$

Here, the effective differential carrier recombination lifetime is a parameter that averages the differential carrier lifetimes in each QW region. Then,

$$\Delta\tau_{rec-eff} = \frac{\sum_{j=1}^{n} \varepsilon_j}{\sum_{j=1}^{n} \varepsilon_j \frac{1}{\Delta\tau_{recj}}} \tag{17}$$

Where,

$$\varepsilon_j = \frac{dN_{wj}}{dN_{w1}} = \frac{(\frac{1}{r_1} + 1)C_{wj}}{(\frac{1}{r_j} + 1)C_{w1}} \tag{18}$$

Similarly, the effective differential carrier escape lifetime is given by:

$$\Delta\tau_{esc-eff} = \frac{\sum_{j=1}^{n} \varepsilon_j}{\sum_{j=1}^{n} \varepsilon_j \frac{r_j}{\Delta\tau_{recj}}} \tag{19}$$

Both $\Delta\tau_{rec-eff}$ and $\Delta\tau_{esc-eff}$ are representative parameters for MQW LEDs using the MCLM and are compatible with the definitions for $\Delta\tau_{rec}$ and $\Delta\tau_{esc}$ in quasi-SQW LEDs using the SCLM.

Due to the large number of independent parameters in the MCLM, fitting experimental data with the model is very difficult using direct fitting algorithms. The increase in independent fitting parameters with more QW regions in the MCLM leads to over definition of the circuit, meaning that multiple equivalent circuits can satisfy the same constrained conditions. In this work, the simultaneous fitting of the impedance and modulation response is based on the gradient descent method [35], an optimization algorithm which finds local minima or maxima. The circuit elements are adjusted using the gradient descent method to achieve the best fit to the real and imaginary parts of the impedance and modulation response. To illustrate the implementation of the MCLM, we start with the simplest case of a two-lifetime model. We start with 50k sets of initial guesses of the circuit element values (i.e., 50k unique circuits) and then search for local best fittings for each circuit by varying the circuit element values by different amounts. We begin with a 10% deviation for each element and subsequently reduce the deviation as the circuit elements converge to local best fittings of the deviation level. The deviation decreases to 2%, 0.4%, 0.1%, and finally settles at 0.03% after the local best fittings at various deviation levels are achieved using the gradient descent method. Here, we use the conventional gradient descent method, which varies one element value at a time, and the fittings stabilize with a deviation level of 0.1% or less. See supplementary material for more details about the convergence. Fittings with $R_{max}^2$ above 0.995 can be achieved over a wide range of current densities. $R_{max}^2$ is defined as the highest $R^2$ value out of the 50k local best fitting circuits. $R^2$ in a given equivalent circuit is taken as the smallest $R^2$ out of the fittings of the $S_{11}(real)$, $S_{11}(imaginary)$, and $S_{21}$. Then, we define the good fittings as a subset of the local best fitting circuits having a fitting quality between $[R_{max}^2 - \Delta R^2, R_{max}^2]$, where $\Delta R^2 = 0.002$. We demonstrate that the choice of the range of the good fittings is robust in the next section.

The quasi-SQW and MQW LEDs described above were analyzed with the SCLM and two-lifetime MCLM, respectively. Cladding layer related equivalent circuit elements are directly associated with the carrier density and are important in understanding the carrier behaviors in the cladding layer. Here, the results for $R_s$ and $C_t$ are shown in Figure 6(a) and 6(b), respectively. In Figure 6 and the subsequent figures, line plots represent the quasi-SQW LEDs, while boxplots represent the MQW LEDs. In the boxplots, the error bars indicate the entire range of the data, the top and bottom of the box represent the first and third quartiles of the data set, and the line inside the box represents the median value of the data set. Due to the better carrier transport in the MQW LED, the carrier density at a given $J$ is expected to be lower. The higher $R_s$ and lower $C_t$ values for the MQW LED confirm this expectation and illustrate a key advantage of MQW LEDs with V-pit engineering. A detailed analysis of $R_s$ and $C_t$ can be found in Ref. 32.

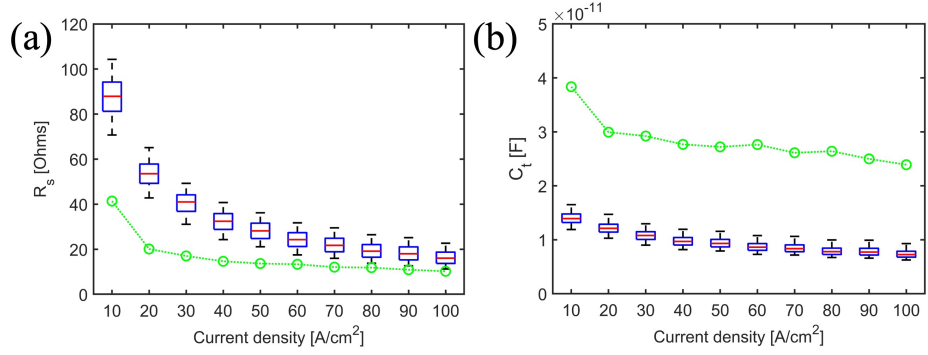

Figure 6. (a) $R_s$ and (b) $C_t$ of quasi-SQW and MQW LEDs. Line plots represent the quasi-SQW LEDs, while boxplots represent the MQW LEDs.

The carrier recombination lifetime plays a crucial role in quantifying the dynamics of carriers in QWs and provides valuable insights into carrier behavior. In the two-lifetime model, we represent two regions with different differential recombination lifetimes as $\Delta\tau_{rec1}$ and $\Delta\tau_{rec2}$, where $\Delta\tau_{rec1} < \Delta\tau_{rec2}$. Figures 7(a) to (c) illustrate the differential recombination lifetime in the quasi-SQW LED ($\Delta\tau_{rec}$), the two constituent differential recombination lifetimes in the MQW LED ($\Delta\tau_{rec1}$ and $\Delta\tau_{rec2}$), and the effective differential recombination lifetime in the MQW LED ($\Delta\tau_{rec-eff}$). Here, the $\Delta\tau_{rec-eff}$ is acquired from $\Delta\tau_{rec1}$ and $\Delta\tau_{rec2}$ using equation (17). In Figures 7(a) to (c), the open dots represent the differential recombination lifetime in the quasi-SQW LED and are identical across the plots. The boxplots in Figures 7(a) to (c) depict $\Delta\tau_{rec1}$, $\Delta\tau_{rec2}$, and $\Delta\tau_{rec-eff}$ in the MQW LED, respectively. Notably, $\Delta\tau_{rec-eff}$ lies between the values of $\Delta\tau_{rec1}$ and $\Delta\tau_{rec2}$. At higher $J$, $\Delta\tau_{rec1}$, $\Delta\tau_{rec2}$, $\Delta\tau_{rec-eff}$, and $\Delta\tau_{rec}$ decrease, indicating faster carrier recombination. This is a consequence of the increased carrier density at higher $J$, leading to higher carrier population and recombination rates. It is worth mentioning that $\Delta\tau_{rec1}$, $\Delta\tau_{rec2}$, and $\Delta\tau_{rec-eff}$ are all higher compared to $\Delta\tau_{rec}$, suggesting that carriers in any region within a V-pit engineered MQW LED recombine at a slower rate compared to the quasi-SQW LED. This observation confirms the benefits of using V-pit engineered MQW LEDs for enhancing the carrier transport in the active region and decreasing the carrier density. In Figure 7(d), the values of $\Delta\tau_{rec-eff}$ are presented with different definitions of good fitting ranging from $\Delta R^2 = 0.008$, $\Delta R^2 = 0.004$, and $\Delta R^2 = 0.002$. Here, the result of $\Delta\tau_{rec-eff}$ is stable over the range of good fitting definitions as the values of $\Delta\tau_{rec-eff}$ are all local best fittings.

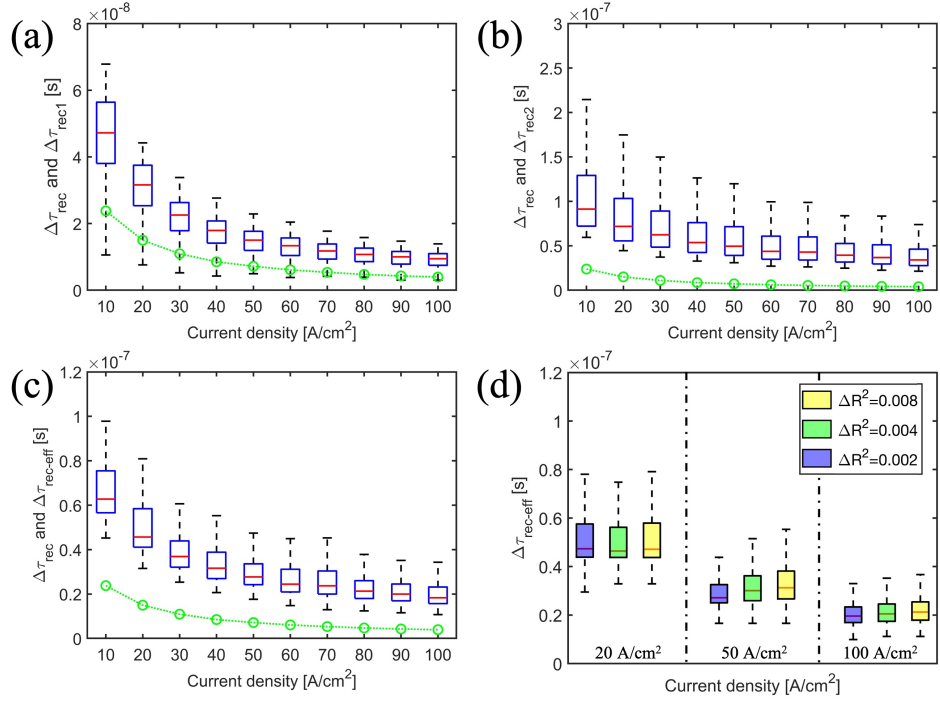

Figure 7. (a) $\Delta\tau_{rec}$ and $\Delta\tau_{rec1}$, (b) $\Delta\tau_{rec}$ and $\Delta\tau_{rec2}$, (c) $\Delta\tau_{rec}$ and $\Delta\tau_{rec-eff}$, and (d) $\Delta\tau_{rec-eff}$ at different fitting quality ranges at 20, 50, and 100 A/cm². Line plots represent the quasi-SQW LEDs, while boxplots represent the MQW LEDs.

Similarly, the effective differential escape lifetime describes the time that it takes for carriers in the QW to escape into the cladding layer and can be acquired from the MCLM. The $\Delta\tau_{esc-eff}$ for the MQW LED and $\Delta\tau_{esc}$ for the quasi-SQW LED are illustrated in Figure 8. It is noted that $\Delta\tau_{esc-eff}$ is more than one order higher than $\Delta\tau_{esc}$, indicating the MQW LED has much better carrier retention in the QWs. $\Delta\tau_{esc-eff}$ is also more than one order higher than $\Delta\tau_{rec-eff}$, showing that most carriers in the MQW LED QWs recombine instead of escape to the cladding layer. Compared with the differential recombination lifetimes, the differential escape lifetimes are more stable as a function of *J*. Recombination in the QWs occurs faster when carriers start to build up in the active region due to recombination mechanisms being positively correlated with carrier density. However, carrier escape is a competing mechanism that is suppressed by increased carrier recombination, and therefore no rate increase is observed.

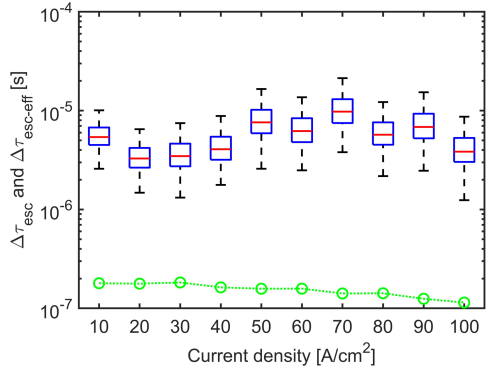

Figure 8. $\Delta\tau_{esc}$ and $\Delta\tau_{esc-eff}$ for quasi-SQW and MQW LEDs. Line plots represent the quasi-SQW LEDs, while boxplots represent the MQW LEDs.

Next, we examine the modulation response in regions characterized by different carrier recombination lifetimes. The following discussion is based on measured data, and the differential carrier lifetimes are obtained from fittings of the MCLM using equation (17). According to the theoretical section presented earlier, the modulation response of MQWs can be analyzed independently for each region with different differential carrier lifetimes. Here, we identify two broad scenarios within the two-lifetime model and analyze them using identical experimental data. The first scenario involves two QW regions, one of which exhibits *faster* recombination but a *lower* light output than the other. An example of this scenario is depicted in Figure 9(a), showing data from an MQW LED operating at 100 A/cm² with two distinct differential carrier lifetimes of 8.45 ns and 29.2 ns. Conversely, the second scenario arises when one QW region exhibits *faster* recombination and a *higher* light output than the other. Figure 9(b) shows this scenario, demonstrating two recombination regions with differential carrier lifetimes of 13.0 ns and 75.1 ns. The effective carrier lifetimes for these two scenarios are 18.0 ns and 23.1 ns, respectively. Based on the available information, it is not possible to ascertain which scenario accurately represents the LED since both combinations produce excellent fittings. However, regardless of the specific scenario, a similar *effective* differential carrier lifetime ($\Delta\tau_{rec-eff}$) is obtained. This implies that the impedance and modulation response characteristics are unique to the specific carrier recombination process, regardless of the interpretation. Therefore, by utilizing information about the impedance and modulation response, it is feasible to predict the $\Delta\tau_{rec-eff}$ accurately. Moreover, we can use the $S_{21}$ roll-off as a predictor of carrier distribution uniformity in MQWs, with a slower roll-off indicating more non-uniformity. At a given frequency, the modulation response is dominated by the QW region with the highest amplitude and is also affected by QW regions with smaller amplitudes.

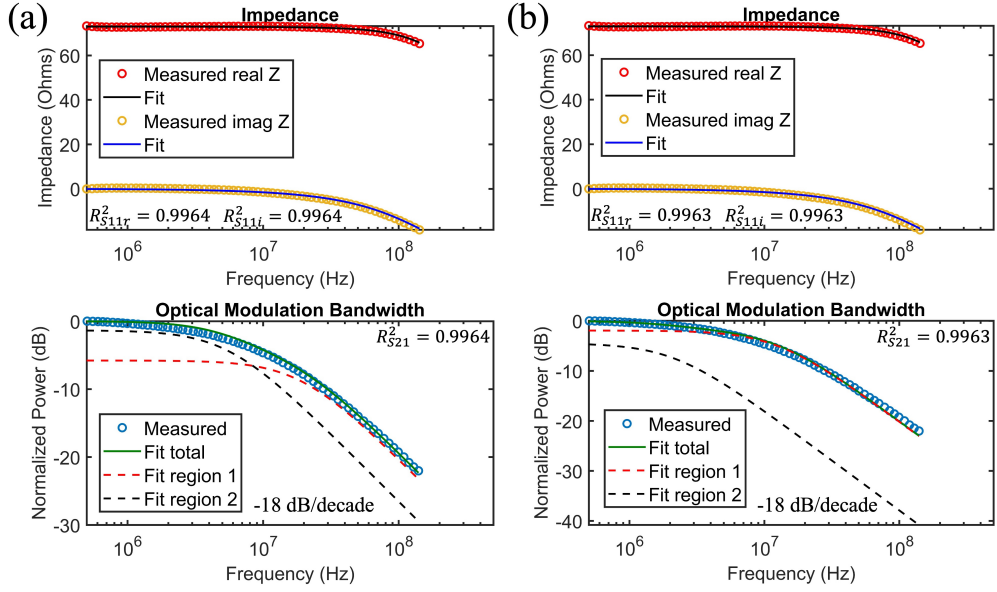

Figure 9. Two different scenarios for MQW LEDs. (a) Faster recombination with a lower light output, (b) Faster recombination with a higher light output.

Since the MCLM is generally applicable for non-uniform carrier distribution in LEDs, and not specifically for MQW LEDs with V-pit engineering, the model can also be used in other types of LEDs with non-uniform lateral carrier distribution. Carrier dynamics studies in micro-LEDs also present a similar non-uniform carrier distribution challenge due to strong surface recombination and the increase of surface-to-volume ratio as the mesa size shrinks [36,37]. The MCLM should be useful in such LEDs too.

Conclusion

In summary, we developed a multiple carrier lifetime SSEL model for MQW InGaN/GaN LEDs with non-uniform carrier distribution. The model is capable of analyzing LEDs with modulation response with $S_{21}$ roll-off of slower than -20 dB/decade. By utilizing this model, we acquired various carrier dynamics parameters associated with carrier transport and recombination, including effective differential carrier escape and recombination lifetimes. Through a comparative analysis between MQW LEDs with V-pit engineering using the MCLM and quasi-SQW LEDs using the SCLM, we observed a slower carrier recombination rate in the MQW LEDs, as well as weak carrier escape from the QWs. These findings confirm the advantages of V-pit engineering in terms of both reducing carrier density and suppressing carrier leakage from the QWs at a given *J*. We validated the MCLM using a three-fold validation process and demonstrated that the model is reliable and robust. Furthermore, the MCLM can be extended and applied to other types of LEDs with non-uniform carrier distribution, such as micro-LEDs.

See the supplementary material for the deviation level and fitting quality of the gradient descent method, the three-fold validation of the MCLM, and the symbolic calculation of impedance and modulation response of the equivalent circuit of the MCLM.

This work was supported by the Department of Energy under Award No. DE-EE0009163.

# Multiple-Carrier-Lifetime Model for Carrier Dynamics in InGaN/GaN LEDs with Non-Uniform Carrier Distribution (Supplementary Material)

Xuefeng Li[1*], Elizabeth DeJong[1], Rob Armitage[2], and Daniel Feezell[1]

[1]*Center for High Technology Materials (CHTM), University of New Mexico, Albuquerque, NM 87106, USA*
[2]*Lumileds LLC, San Jose, CA 95131, USA*

***Electronic mail:** xuefengli@unm.edu

**Deviation level and fitting quality of gradient descent method**

$R^2$ in a given equivalent circuit is taken as the smallest $R^2$ out of the fittings of the $S_{11}(real)$, $S_{11}(imaginary)$, and $S_{21}$. We present the $R^2$ values in the local best fitting circuits after applying a gradient descent process at various deviation levels, ranging from 10% to 0.03%. The results are shown in Figure S1. It is shown that as the deviation level decreases, the fitting quality improves, eventually stabilizing with a deviation level of 0.1% or less. Here, the *x* axis represents the ranking of fitting quality among all the circuits at a given deviation level. The higher the ranking index, the better the fit.

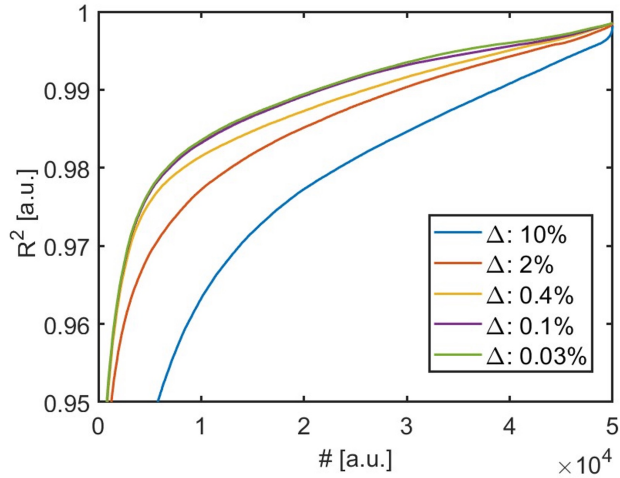

Figure S1. Deviation level and fitting quality of gradient descent method.

**Validation of the multiple carrier lifetime model**

The presented multiple carrier lifetime model offers important insights into the effective differential carrier lifetime. To assess the reliability of this model, we present a three-fold validation process. Firstly, we demonstrate that the gradient descent process utilized for the two-lifetime model can consistently replicate the effective differential carrier lifetime, even with different initial circuit element values. Figure S2(a) depicts the effective differential carrier lifetime obtained by employing another 50k random alternative initial guesses for the circuit elements using the same procedure and shows indistinguishable results to those shown in Figure 7(c). This indicates that the chosen number of initial guesses (i.e., 50k in this case) is sufficiently large to yield a stable outcome.

Another validation perspective we explore is the robustness of the MCLM in a three-carrier lifetime model. Theoretically, this multiple carrier lifetime approach can accommodate an infinite number of carrier lifetimes, with greater accuracy achieved as more lifetimes are included. However, a larger number of lifetimes introduce additional parameters, which makes the procedure computationally expensive and a robust fitting hard to acquire. Figure S2(b) shows the effective differential carrier lifetime ($\Delta\tau_{rec-eff}$) obtained using a three-carrier lifetime version of the MCLM. The fact that the effective lifetimes from the two-carrier and three-carrier lifetime models match so well suggests that two carrier lifetimes are sufficient for a reliable analysis of carrier dynamics in these particular MQW LEDs.

Additionally, we investigate the compatibility of the MCLM and the SCLM both applied to the quasi-SQW LEDs emitting from a single QW region. For example, we apply the new MCLM model to the quasi-SQW LEDs and compare the results. We can understand the results by considering a homogeneous LED as having multiple identical regions, each with the same carrier lifetime. Consequently, the effective differential carrier lifetime ($\Delta\tau_{rec-eff}$) in the MCLM should equate to the recombination lifetime ($\Delta\tau_{rec}$) in the SCLM. Figure S2(c) illustrates the $\Delta\tau_{rec}$ values obtained using the SCLM (green line) and the MCLM (box plots) for the quasi-SQW LED. The agreement of results between the two models indicates that the MCLM offers a more general representation of the carrier dynamics in LEDs, i.e., the SCLM has identical carrier lifetime across the entire active region and is a special case of the MCLM. Through these three validation processes, we establish the robustness and reliability of the proposed model.

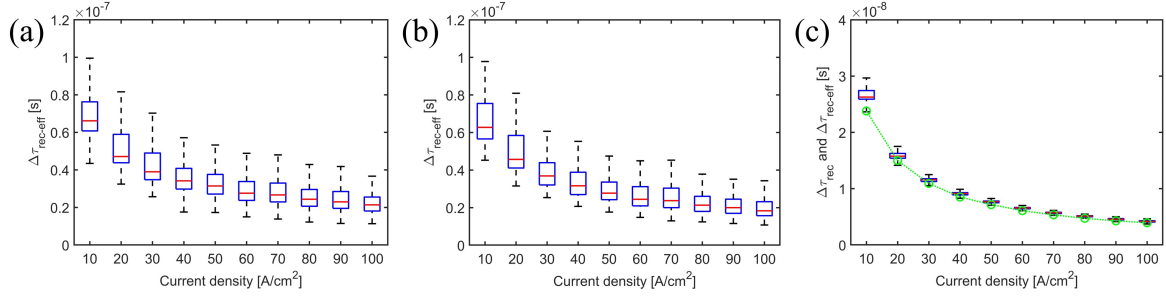

Figure S2. Validation of the MCLM. (a) Replication of the two-carrier lifetime model with new initial conditions, (b) $\Delta\tau_{rec-eff}$ from a three-carrier-lifetime model, (c) Differential carrier lifetimes for quasi-SQW LED using SCLM (green line) and two-carrier-lifetime model (box plots).

**Symbolic calculation of impedance and modulation response**

The impedance of the equivalent circuit in the MCLM (Figure 5 in the paper) can be acquired using symbolic calculation, and the corresponding real and imaginary parts of the impedance are:

$$Z_{real} = \frac{a}{\omega^2 b^2 + a^2} + R_s \tag{1}$$

$$Z_{imag} = \frac{-\omega b}{\omega^2 b^2 + a^2} \tag{2}$$

Where, the *a* and *b* terms are as follows:

$$a = \sum_{j=1}^{n} \frac{r_j}{R_{wj}} \left[ \frac{\Delta\tau_{recj}^2 \omega^2 + r_j + 1}{\left(1 + r_j\right)^2 + \Delta\tau_{recj}^2 \omega^2} \right] \tag{3}$$

$$b = C_t + \sum_{j=1}^{n} \frac{\Delta\tau_{recj}^2}{R_{wj}} \left[ \frac{r_j^2}{\left(1 + r_j\right)^2 + \Delta\tau_{recj}^2 \omega^2} \right] \tag{4}$$

A similar method can also be used to calculate the modulation response and an accurate expression for QW region *j*, which was discussed in equation (15) in the paper, is:

$$\left|\frac{V_{out}}{V_{in}}\right|^2 = \frac{r_j}{R_{wj}(r_j + 1)} * \frac{c}{(a^2 + b^2\omega^2)\left(\left(r_j + a/(a^2 + \omega^2 b)\right)^2 + b^2\omega^2/(a^2 + \omega^2 b)^2\right)} \tag{5}$$

Where $c = \frac{r_j^2}{(r_j+1)^2 + \Delta\tau_{recj}^2 \omega^2}$. Then the denominator term in equation (5) is related to $\left|\frac{Z_{tot}-R_S}{Z_{tot}}\right|^2$ from equation (15) in the paper.